\documentstyle[preprint,epsfig,aps]{revtex}

\begin{document}
\title{Density dependence of in-medium nucleon-nucleon cross sections}
\author{C. A. Bertulani}
\address{Department of Physics, Brookhaven National Laboratory,\\
Upton, NY 11973-5000, USA}
\maketitle

\begin{abstract}
The lowest-order correction of the 
density dependence of  in-medium nucleon-nucleon
cross sections is obtained from geometrical
considerations  of Pauli-blocking effects. As a by-product,
it is shown that the medium corrections imply an $1/E$ 
energy dependence of the density-dependent term. 

\end{abstract}



\vskip 2cm

The nucleon-nucleon cross section is a fundamental input in theoretical
calculations of nucleus-nucleus collisions at intermediate and high  
energies ($E/A \gtrsim 100$ MeV). One expects to obtain information
about the nuclear equation of state by studying global collective
variables in such collisions (see, e.g., \cite{Gut89}).  
Transport equations, like the BUU equation, are often used as tools 
for the analysis of experimental data and as a
bridge to the information about the equation of state 
(see, e.g., \cite{Wes93}).
The nucleon-nucleon cross sections are building blocks in these
transport equations.

In previous theoretical studies of 
heavy ion collisions at intermediate energies ($E/A \simeq 100$ MeV)
the nucleon-nucleon cross
section was multiplied with a constant scaling factor to account for in-medium
corrections \cite{BLT87,WB88}. As pointed out in ref. \cite{Wes93}, this
approach fails in low density nuclear matter where the in-medium
cross section should approach its free-space value. A more realistic
approach uses a Taylor expansion of the in-medium cross section
in the density variable. One gets \cite{KWB93}
\begin{equation} 
\sigma_{NN}  = \sigma_{NN}^{free} \left( 1+ \alpha
{\bar \rho} \right)
 \ , 
\label{sig} 
\end{equation} 
where ${\bar \rho}=\rho/\rho_0$, $\rho_0$ is the normal nuclear density,
and $\alpha$ is the logarithmic derivative of the in-medium cross section
with respect to the density, taken at $\rho =0$, 
\begin{equation} 
\alpha = \rho_0 {\partial \over \partial \rho} \left(
\ln \sigma_{NN}\right)\mid_{\rho=0}
 \ . 
\label{sig2} 
\end{equation} 
This parameterization is motivated by Br\"uckner G-matrix theory and is
basically due to Pauli-blocking of the cross section for 
collisions at intermediate energies \cite{Sch98}.     
Values of $\alpha$ between $-0.4$ and $-0.2$ yield the best agreement
with involved G-matrix calculations using realistic nucleon-nucleon
interactions \cite{Sch98}.

In this article we give a simple and transparent derivation of the lowest 
order expansion of the in-medium nucleon-nucleon cross section in terms
of the nucleon density. In our approach the leading term of the expansion 
comes out as $\alpha' \rho^{2/3}$ with  $\alpha'$  proportional
to $1/E$.  
This energy-dependence 
agrees with  experimental results on total nucleus-nucleus
cross sections.

We adopt the idea that the main effect of medium corrections is due to
the Pauli-blocking of nucleon-nucleon scattering.  Pauli-blocking prevents
the nucleons to scatter into final occupied states in binary collisions
between the projectile and target nucleons. This is best seen in momentum
space, as shown in figure 1. We see that energy and momentum
conservation, together with the Pauli principle, restrict the collision phase
space  to a 
complex geometry involving the Fermi-spheres and the scattering sphere.   
In this scenario, the
in-medium cross section corrected by Pauli-blocking can be defined
as
\begin{equation}
\sigma_{NN} (k, K_{F1}, K_{F2}) = \int {d^3k_1d^3k_2 \over (4\pi
K_{F1}^3/3)(4\pi K_{F2}^3/3)} \ {2q\over k} \ \sigma_{NN}^{free} (q) \ \ 
{\Omega_{Pauli} \over 4\pi} \ ,
\label{ave}
\end{equation}
where $k$
is the relative momentum  per nucleon of the nucleus-nucleus collision
(see figure 1), and $\sigma_{NN}^{free} (q)$ is the free nucleon-nucleon cross
section for the relative momentum $2{\bf q} =  {\bf k}_1 - {\bf k}_2 -{\bf
k}$, of a given pair of colliding nucleons. Clearly, Pauli-blocking enters
through the restriction that $|{\bf k'}_1|$ and $|{\bf k'}_2|$ lie outside the
Fermi spheres. From energy and momentum conservation in the collision, ${\bf
q'}$ is a vector which can only rotate around a circle with
center at ${\bf p} = ( {\bf k}_1 - {\bf k}_2 -{\bf
k})/2$.  These conditions yield an allowed scattering  solid angle given by \cite{Ber86}
\begin{equation}
 \Omega_{Pauli} =  4\pi - 2( \Omega_a + \Omega_b - {\bar \Omega})
\ ,
\label{pauli}
\end{equation}
where $\Omega_a$ and $\Omega_b$ specify the excluded
solid angles for each nucleon, and $\bar \Omega$ represents the intersection
angle of $\Omega_a$ and $\Omega_b$ (see figure 1).

The solid angles $\Omega_a$ and $\Omega_b$ are easily determined. They are given by
\begin{equation}
\Omega_a = 2 \pi (1-\cos \theta_a ) \ , \ \ \  \ \ \ \
\Omega_b = 2 \pi (1-\cos
\theta_b ) \ ,
\label{Omega}
\end{equation}
where  $\bf q$ and $\bf p$ were defined above, ${\bf b}= {\bf k} - {\bf
p}$, and
\begin{equation}
\cos \theta_a = (p^2+q^2-K_{F1}^2)/2pq \ , \ \ \  \ \ \ \
\cos \theta_b = (p^2+q^2-K_{F2}^2)/2pq \ ,
\label{thetaab}
\end{equation}

The
evaluation of  $\bar \Omega$ is tedious but can be done analytically.
The full calculation was  done in ref. \cite{Ber86} and 
the results have
been reproduced in the appendix of ref. \cite{HRB91}. To summarize,
there are two possibilities:
\begin{eqnarray}
(1) \ \ \ \ \ {\bar \Omega} &=& \Omega_i (\theta, \theta_a,
\theta_b) +  \Omega_i(\pi-\theta,
\theta_a, \theta_b) \ , \ \ \ \ {\rm for} \ \
\theta+\theta_a+\theta_b > \pi \\
(2) \ \ \ \ \ {\bar \Omega} &=& \Omega_i (\theta, \theta_a,
\theta_b)\ ,  \ \ \ \ {\rm for} \ \
\theta+\theta_a+\theta_b \leq \pi  \ ,
\label{O1}
\end{eqnarray}
where $\theta$ is given by
\begin{equation}
\cos \theta = (k^2-p^2-b^2)/2pb \ .
\label{O2}
\end{equation}
The solid angle $\Omega_i$ has the following values
\begin{eqnarray}
(a) \ \ \ \ \ \Omega_i &=& 0\ , \ \ \ {\rm for} \ \
\theta\geq \theta_a+\theta_b  \\
(b)\ \ \ \ \   
\Omega_i &=& 
2 \left[ \cos^{-1} \left(
{\cos \theta_b - \cos \theta \cos \theta_a \over
\sin\theta_a ( \cos^2\theta_a +\cos^2\theta_b - 
2 \cos\theta \cos\theta_a\cos\theta_b)^{1/2}}\right)\right. 
\nonumber \\
&+&
\cos^{-1} \left(
{\cos \theta_a - \cos \theta \cos \theta_b \over
\sin\theta_b ( \cos^2\theta_a +\cos^2\theta_b - 
2 \cos\theta \cos\theta_a\cos\theta_b)^{1/2}}\right)
\nonumber \\
&-&
\cos\theta_a \cos^{-1} \left(
{\cos \theta_b - \cos \theta \cos \theta_a \over
\sin\theta \sin\theta_a}\right)
\nonumber \\
&-&
\left.
\cos\theta_b \cos^{-1} \left(
{\cos \theta_a - \cos \theta \cos \theta_b \over
\sin\theta \sin\theta_b}\right)\right]\\
&{\rm for}& \ \ \
|\theta_b - \theta_a | \leq \theta \leq 
\theta_a+\theta_b \ ,\\
(c) \ \ \ \ \ \Omega_i &=& \Omega_b \ \ \ {\rm for} \ \ \
\theta_b  \leq \theta_a , \  
\theta\leq |\theta_b - \theta_a| \ ,\\
(d) \ \ \ \ \ \Omega_i &=& \Omega_a \ \ \ {\rm for} \ \ \
\theta_a  \leq \theta_b , \  
\theta\leq |\theta_b - \theta_a| \ ,
\label{big}
\end{eqnarray}

The integrals over ${\bf
k}_1$ and ${\bf k}_2$ in (\ref{ave}) reduce to a five-fold integral due to
cylindrical symmetry. Two approximations can be done which greatly
simplify  the problem: (a) on average, the symmetric situation
in which $K_{F1}=K_{F2}\equiv K_F$, ${\bf q} = {\bf k}/2$, ${\bf p} = {\bf k}/2$, and
${\bf b} = {\bf k}/2$, is favored, (b) the  free nucleon-nucleon cross section
can be taken outside of the integral in eq. (\ref{ave}).
Both approximations are supported by the
studies of refs. \cite{HRB91} and can be  verified 
numerically
\cite{Ber86}. The assumption (a) implies that $\Omega_a=\Omega_b={\bar
\Omega}$, which can be checked using the equation (\ref{big}). 
One gets from (\ref{pauli}) the simple
expression 
\begin{equation}
 \Omega_{Pauli} =  4\pi - 2 \Omega_a =4\pi\left( 1- 2 {K_F^2
\over k^2}\right) 
\ .  \label{pauli2}
\end{equation}
Furthermore, the assumption (b) implies that
\begin{equation}
\sigma_{NN} (k, K_F) =  \sigma_{NN}^{free} (k)   
\ {\Omega_{Pauli} \over 4\pi} =  \sigma_{NN}^{free} (k)\  \left( 1- 2 {K_F^2
\over k^2}\right) \ .
\label{ave2} 
\end{equation}
 
The above equation shows that the in-medium
nucleon-nucleon cross section is about
1/2 of its free value for $k=2K_F$, i.e., for $E/A \simeq 150$ MeV,
in agreement with the numerical results of ref. \cite{HRB91}.
 
The connection  with the nuclear densities
is accomplished through the local density approximation, which
relates the Fermi momenta to the local densities as
\begin{equation}
K_{F}^2 = \left[ {3 \pi^2 \over 4} \rho(r) \right]^{2/3} + 
{5\over 2} \xi \left( \nabla \rho / \rho \right)^2 \ , \label{KF}
\end {equation}
where $\rho(r)$ is the sum of the nucleon densities of each colliding
nucleus at the position $r$.
The second term is small and amounts to a surface correction, with $\xi$ of the
order of 0.1 \cite{HRB91}. 

Inserting (\ref{KF}) into (\ref{ave2}), and using $E=\hbar^2 k^2/2m_N$, we get
(with ${\bar \rho}=\rho/\rho_0$)
\begin{equation}
\sigma_{NN} (E, \rho) =  \sigma_{NN}^{free} (E)   
 \left(1+ \alpha' {\bar \rho}^{2/3}\right) \ \ \ \ \ {\rm where}
\ \ \ \alpha'=-{48.4\over E\ (MeV)}
\label{ave3} 
\end{equation}
where the second term of (\ref{KF}) has been neglected. This equation shows that
the local density approximation 
leads
to a density dependence  proportional to
${\bar \rho}^{2/3}$. The Pauli principle
yields a  $1/E$ dependence on the bombarding energy. This behavior 
arises from  a larger phase space  available for nucleon-nucleon scattering
with increasing energy. 

The nucleon-nucleon cross section at $E/A \lesssim 200$ MeV
decreases with $E$ approximately as $1/E$. We thus expect that, 
in nucleus-nucleus collisions,
this energy
dependence is flattened  by the Pauli correction,
i.e., the in-medium nucleon-nucleon cross section  is more flat as a function
of $E$, for $E \lesssim 200$, than the free cross section. 
For higher values of $E$ the Pauli blocking is
less important and the free and in-medium nucleon-nucleon cross sections are
approximately equal. These conclusions are in agreement with the experimental
data for nucleus-nucleus
reaction cross sections \cite{Kox85}. This was in fact well explained in
ref. \cite{HRB91}. 

Notice that, for $E/A = 100-200$ MeV, and  $\rho \simeq \rho_0$, eq. (\ref{ave3}) yields a coefficient 
$\alpha'$ between $ -0.2$ and $ - 0.5$. 
This is  in excellent agreement with the findings based on the BUU 
calculations, primarily intended to reproduce the experimental data on 
collective variables in intermediate energy nucleus-nucleus collisions.

In conclusion, we have presented a microscopic derivation of the 
lowest order density correction for the in-medium nucleon-nucleon cross
section. Despite its simplicity, the calculation shows that 
Pauli-blocking is able to explain almost entirely the magnitude of the
correction term, although the power of density dependent term is slightly
different from what is commonly mentioned in the literature
\cite{Wes93,BLT87,WB88,KWB93,Sch98}.  
We also predict an energy dependence of the in-medium
cross sections which was not accounted for previously. This calls for a
further study of the consequence of this energy dependence in the 
transport equation analysis of collective variables in nucleus-nucleus
collisions at intermediate energies.

\vskip 0.5 cm
{\noindent {\bf Acknowledgments}
\vskip 0.2 cm

The author is a fellow of the John Simon Guggenheim foundation. 
This work has been authored under Contract No. DE-AC02-98CH10886 with the U.S. Department of Energy, and
partial
support from the Brazilian funding agency 
MCT/FINEP/CNPQ(PRONEX), under contract No. 41.96.0886.00, is also acknowledged.

\vskip 0.5 cm

{\noindent \bf Figure Captions}
\vskip 0.2 cm

\begin{enumerate}
\item  
The geometric  description 
of Pauli blocking, in momentum space, for binary collisions
of target and projectile nucleons.

\end{enumerate}

\epsfig{file=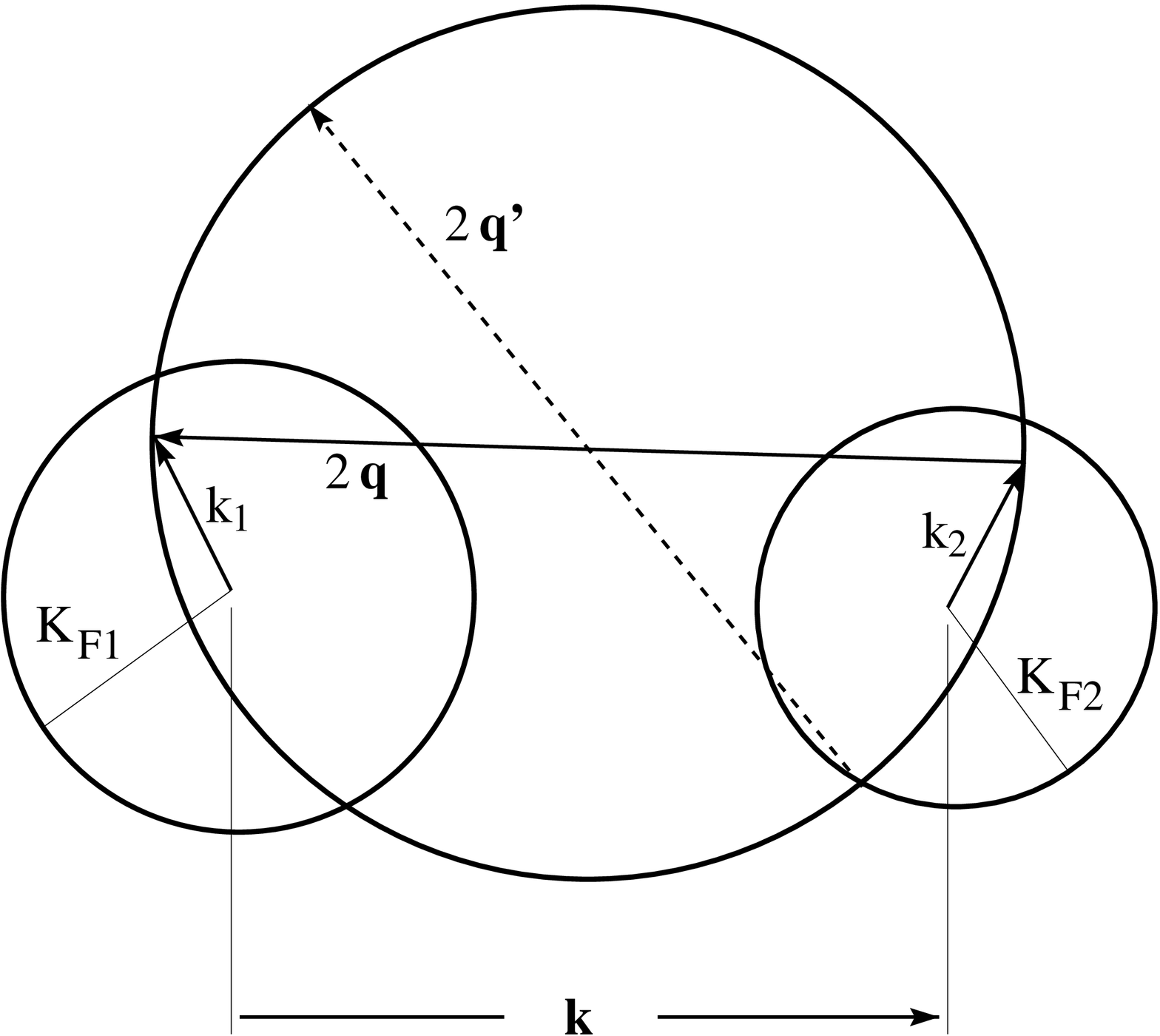,width=.75\textwidth}

\vskip 2cm

\epsfig{file=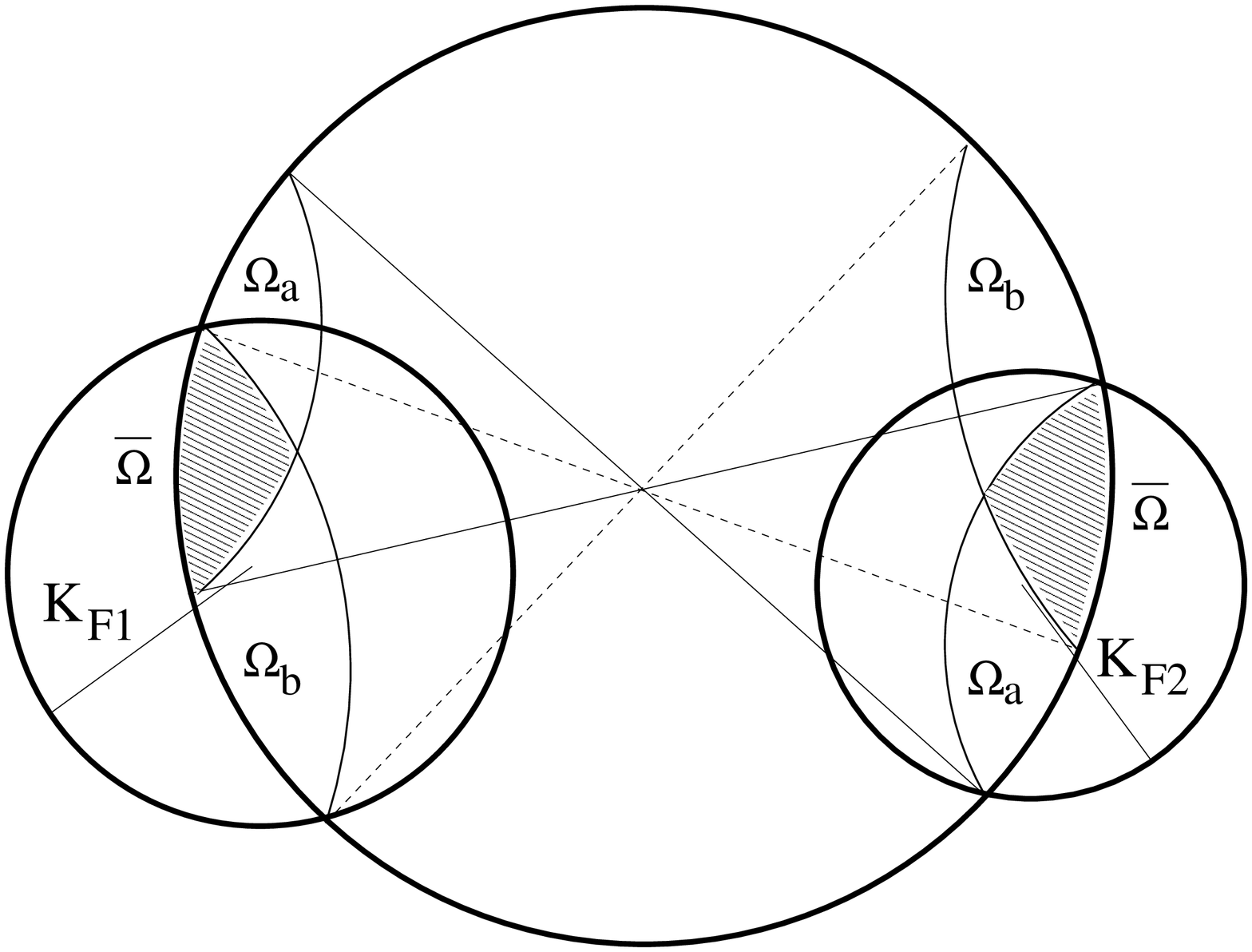,width=.75\textwidth}

\end{document}